\documentstyle[12pt,epsf]{article}
\begin{document}
\setlength{\baselineskip}{0.27in}

%%definitions%%%%%%%%%%%%%%%%%%%%%%%%%%%%%%%%%%%%%%%%%%%%%%%%%%%%%%%%%

\newcommand{\beq}{\begin{equation}}
\newcommand{\eeq}{\end{equation}}
\newcommand{\beqa}{\begin{eqnarray}}
\newcommand{\eeqa}{\end{eqnarray}}
\newcommand{\lsim}{\begin{array}{c}\,\sim\vspace{-21pt}\\<
\end{array}}
\newcommand{\gsim}{\begin{array}{c}\sim\vspace{-21pt}\\> \end{array}}
\newcommand{\nt}{\nu_\tau}
\newcommand{\nee}{\nu_e}
\newcommand{\nm}{\nu_\mu}
\newcommand{\bi}{\bibitem}
%%%%%%%%%%%%%%%%%%%%%%%%%%%%%%%%%%%%%%%%%%%%%%%%%%%%%%%%%%%%%%%%%%%%%%

\begin{titlepage}
\pagestyle{empty}
\rightline{UM-TH-95-23}
\rightline{hep-ph/9509375}
\rightline{September 1995}
\vspace*{0.5cm}
\begin{center}
\vglue .06in
{\Large \bf  $\psi^\prime$~Polarization as a test of colour octet \\
quarkonium production} \\[0.8in]

{\sc M.~Beneke\footnote{
Address after Oct, 1, 1995: SLAC,
P.O.~Box 4349, Stanford, CA~94309, U.S.A.
}}
and
{\sc I.Z.~Rothstein\footnote{
Address after Oct, 1, 1995: University of California, San Diego,
9500 Gilman Drive, La Jolla, CA~92093, U.S.A.
}}

\vskip 1truecm

{\it Randall Laboratory of Physics,} \\
{\it University of Michigan, Ann Arbor, MI 48109-1120, U.S.A.}\\
\vskip 1truecm
\vspace*{0.5cm}

{\bf Abstract}\\[-.05in]

\begin{quote}
We calculate the $\alpha_s$ corrections to the transverse polarization
fraction of $\psi^\prime$'s produced at the Tevatron. If the
`$\psi^\prime$-anomaly' is explained by gluon fragmentation into a
colour octet Fock component of the $\psi^\prime$, the $\psi^\prime$'s
should be $100\%$ transversely polarized at leading order in $\alpha_s$,
up to spin symmetry breaking long-distance corrections.
We find that the short-distance correction to the transverse polarization
fraction is a few percent, so that a polarization measurement would
provide  a reliable test of the colour octet mechanism.
\end{quote}
\end{center}

\end{titlepage}
\setcounter{footnote}{0}

\newpage
\noindent
{\bf 1. Introduction}\\[0.3cm]
\indent
The topic of  quarkonium production has received renewed attention
as of late due to a plethora of data and new theoretical developments.
While it has been known for quite some time that quarkonia calculations are
tractable as a consequence of the fact that the non-perturbative dynamics
may be swept into non-relativistic wave functions, an organizational
principle was still lacking. This shortcoming manifested itself in the fact
that the theory of $P$ wave decays was plagued by infrared divergences that
could not be factorized. This problem was resolved \cite{BOD92} once the
calculation was organized using non-relativistic QCD (NRQCD) \cite{CAS86}.
Within the framework of this effective field theory, the scales involved
in quarkonia calculations are cleanly separated and factorization becomes
explicit.

NRQCD has been widely used in calculations of quarkonium production rates.
For transverse momentum $p_T$ of the quarkonium
larger than the quarkonium mass, the dominant
production mechanism will be fragmentation. This is essentially a resonance
effect resulting from the fact that the off-shellness of the fragmenting
parton is small ($\sim 4 m_c^2$) compared to the transverse momentum
squared. The
rate for this process is enhanced by a factor of $p_T^2/(4m_c^2)$, which
can overcome the suppression due to any additional powers of $\alpha_s$
present in the  fragmentation amplitudes.
The fragmentation functions for quarkonia are partially
calculable in perturbation theory \cite{BRA93}.

If one does not include the fragmentation contribution, the prediction for
the prompt production cross sections for both $J/\psi$ and $\psi^\prime$
at the Tevatron are off by several orders of magnitude.
The inclusion of fragmentation into color singlet quark-antiquark pairs
brings the $J/\psi$ production rate into agreement with experiment within
theoretical errors \cite{CAC94,bdfm,ROY95}, but still leaves the theory
an order of magnitude short for $\psi^{\prime}$ production \cite{bdfm}. A
possible explanation \cite{bf} for this discrepancy is that the dominant
contribution to $\psi^{\prime}$ production results from fragmentation into
two charmed quarks in a relative color octet state.\footnote{Recent
measurement of the $J/\psi$ production fraction from $\chi_c$ decays
indicates that colour octet production is also important to account for
the fraction of $J/\psi$'s not coming from $B$ and $\chi_c$ decays
\cite{cl,CAC95}.} This is easily seen once we consider the process within
NRQCD. In this formalism the full Fock state decomposition of the
$\psi^\prime$,
\beq
\label{fock}
|\psi^\prime\rangle = |c\bar{c}[^3\!S^{(1)}_1]\rangle
+ O(v)\,|c\bar{c}[^3\!P^{(8)}_J] g\rangle
+ O(v^2)\,|c\bar{c}[^1\!S^{(8)}_0]g\rangle
+ O(v^2)\,|c\bar{c}[^3\!S^{(1,8)}_1] gg\rangle
+ \ldots ,
\eeq
is taken into account by the set of all possible production matrix elements
\cite{bbl}. The notation here is spectroscopic $^{2S+1}\!L_J$, with an
additional superscript in parenthesis to denote the colour state.  Each
Fock component is weighted by some power of the heavy quark velocity. This
weight is determined by the velocity scaling rules which will be discussed
below.  Although the transition of a $c\bar{c}$ state in a relative
colour octet state into a $\psi^\prime$ is suppressed by $v^4 \sim (0.3)^2$
relative to the transition from a colour singlet pair, this suppression
is off-set by an $(\pi/\alpha_s(2 m_c))^2$-enhancement for the production
rate of $c\bar{c}$ pairs in a $^3\!S^{(8)}_1$ state. Additional numerical
factors ensure that the value of the matrix element (associated with the
hadronization of the colour octet pair) required for agreement with
experiment is indeed consistent with the velocity scaling
rules \cite{bf}.

As noted by Cho and Wise \cite{cw}, the colour octet
production mechanism could be checked by measuring the polarization of the
$\psi^{\prime}$, as they should be strongly transversely polarized at
leading order in $\alpha_s$. The reason for this asymmetry
is that the fragmenting
gluons are nearly on-shell, so that the $c\bar{c}$ pair inherits the
gluon's transverse polarization up to corrections of
order $(4m_c^2)/q_0^2$, where $q_0$ is the gluon energy in the lab frame. Due
to spin symmetry of the leading order NRQCD Lagrangian, the polarization
of the quark pair stays intact in the subsequent (nonperturbative)
evolution into a $\psi^\prime$ state up to spin symmetry breaking
corrections of order $v^4/3\sim (3-4)\%$.
In this paper we address the question
whether this signature persists after inclusion of radiative corrections
to the short-distance production process and calculate the
order $\alpha_s$ corrections to the prediction of $100\%$
polarization (neglecting spin symmetry breaking).  We find that these
corrections are small, and thus the colour octet proposal can
be easily tested since the polarization will not be destroyed by radiative
corrections. Other deviations from pure transverse
polarization from
(spin-symmetry breaking) long-distance effects or higher twist effects
will most likely be smaller than the short-distance correction, except,
possibly, at small $p_T$, where higher twist effects can become large.\\

\noindent
{\bf 2. Fragmentation functions}\\[0.3cm]
\indent
The fragmentation of a gluon (or any parton) into a quarkonium
is a two-scale process. Over distances of order $1/m_c$ the gluon
fragments into a $c\bar{c}$ pair with longitudinal momentum fraction
$z$ and small relative velocity $v$ in the rest frame of the pair. The
subsequent arrangement of the quark pair into a quarkonium bound state
takes place over longer distances, of order $1/(m_c v)$. To the extent
that $v^2\ll 1$, the effects on both scales can be separated in NRQCD and
the longitudinal momentum fraction of the quarkonium identified with
that of the quark pair, so that the $z$-dependence is perturbatively
calculable. The fragmentation function, defined for instance
as in \cite{COL82}, can then be written as
\beq
\label{factorization}
D_{g\rightarrow H_{\lambda}}(z,\mu)=\sum_n d_{n;ij}(z,\mu,\mu') \,
\langle {\cal O}_{n;ij}^{H_\lambda} \rangle(\mu')
\eeq
where $H_\lambda$ denotes the quarkonium in a specific polarization state
$\lambda$. The $d_{n;ij}$ are short-distance coefficients,
independent of the
quarkonium state and computable as expansions in $\alpha_s$. All
bound state information is summarized in the vacuum matrix elements
of the operators ${\cal O}_{n;ij}^{H_\lambda}$. These parameters
must be determined
phenomenologically or from lattice calculations. Following the notation of
\cite{bbl}, the generic form for the production operators is
\beq
\label{form}
{\cal O}_{n;ij}^{H_\lambda}=
\chi^\dagger \kappa_{n;i}\,\psi\,{a_H^{(\lambda)}}^\dagger
a_H^{(\lambda)}\,\psi^\dagger\kappa^\prime_{n;j}\,\chi ,
\eeq
where $\psi$, $\chi$ are nonrelativistic quark and antiquark fields and
$a_H^{(\lambda)}$ destroys a quarkonium state $H$ with polarization
$\lambda$ in the out-state.
The matrix elements are supposed to be evaluated in the
quarkonium rest frame. The kernels $\kappa_{n;i}$, $\kappa^{\prime}_{n;j}$
contain colour matrices, spin matrices and derivatives. The colour and
spin indices are suppressed. The operators ${\cal O}_{n;ij}^{H_\lambda}$
depend on two sets of three-vector indices $i$ and $j$. If we
consider unpolarized production and include summation
over $\lambda$ in the definition (\ref{form}), the matrix
elements $\langle {\cal O}_{n;ij}^H\rangle$ can not depend on any
three-vector. A tensor decomposition then reduces the sum in
(\ref{factorization}) to a sum of a few scalar operators as listed in
\cite{bbl}. In the case of polarization this decomposition also includes
the polarization tensor of the quarkonium. After decoupling of indices
from the short-distance amplitude, the kernels $\kappa_n$
contain projections on the spin and angular momentum orientation
of the quark pair.
The dependence on the factorization scale for NRQCD,
$m_c v < \mu' < m_c$, drops out in the sum (\ref{factorization}),
while the dependence on
$\mu$, which separates the gluon production from the fragmentation
process is governed by the usual Altarelli-Parisi evolution.\\

\begin{figure}[t]
   \vspace{0cm}
   \epsfysize=10cm
   \epsfxsize=10cm
   \centerline{\epsffile{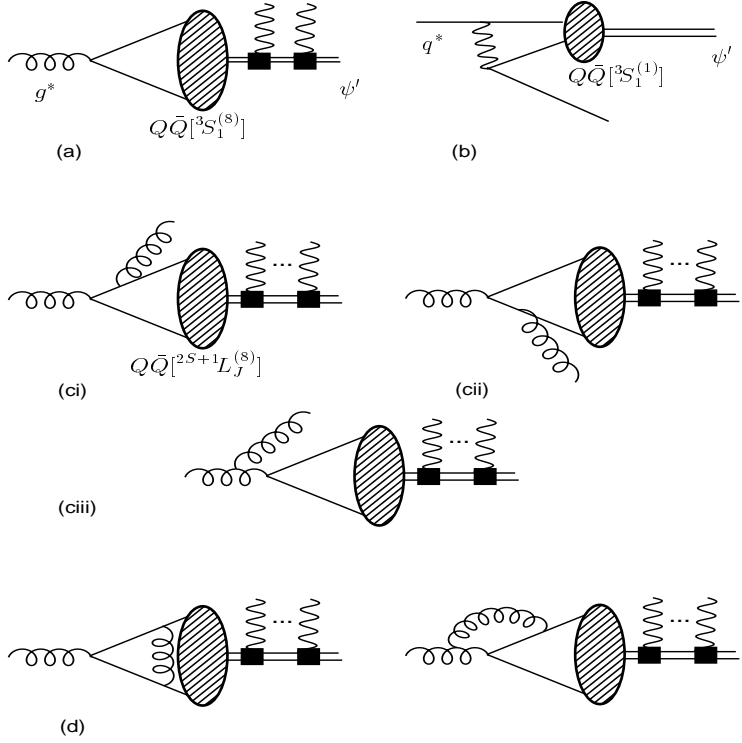}}
   \vspace*{0cm}
\caption{\label{diagplot} (a) Leading order gluon fragmentation into
$c\bar{c}[^3\!S^{(8)}_1]$. (b) Charm fragmentation. (c), (d) Real and
virtual fragmentation diagrams at order $\alpha_s^2$. The black boxes
denote nonperturbative evolution of the $c\bar{c}$-pair into a
$\psi^\prime$ through soft gluon emission.}
\end{figure}

\noindent
{\bf 3. Colour octet production at leading order}\\[0.3cm]
\indent
Given the Fock state decomposition (\ref{fock}), the leading contribution
in $v^2$ comes from gluon and charm fragmentation into the colour
singlet $^3\!S_1$ state. Charm fragmentation is shown\footnote{This diagram
as well as all others are understood to be calculated in axial gauge.
Otherwise, diagrams with gluons attached to eikonal lines would also
have to be considered, see \cite{COL82}.} in
Fig.~\ref{diagplot}b. Everything prior to the shaded oval is considered
as short-distance dominated, whatever is to the right is contained
in the matrix element of an operator ${\cal O}_n^H$.
In Fig.~\ref{diagplot}b, this is simply the non-relativistic wave-function
(squared) at the origin. Using the velocity scaling rules derived in
\cite{LEP92}, colour singlet charm and gluon fragmentation yield
contributions of order $\alpha_s^2v^3$ and $\alpha_s^3 v^3$ respectively.
The net contribution from singlet
fragmentation falls short of the experimental data by a factor of about
30 \cite{bdfm}. However, since $v^2\sim 0.3$ is not small, the contribution
from operators, whose matrix elements are suppressed in $v^2$,
may be important. This is especially so for ${\cal O}_8^{\psi^\prime}
(^3\!S_1)$, because it is the only operator with a short-distance
coefficient $d(z)$ at order $\alpha_s$. A quark pair in a colour octet
$^3\!S_1$ state has non-zero overlap with a Fock state component
of the $\psi^\prime$ that contains two additional gluons, represented
by the emission from the black boxes in Fig.~\ref{diagplot}a. The
long-distance dynamics is accurately described by the NRQCD Lagrangian
and the velocity scaling rules apply. The quark pair in the
$^3\!S_1^{(8)}$ state can evolve into the $\psi^\prime$ by a double
electric dipole transition, each of which contributes a factor of
$v$ to the amplitude as indicated in (\ref{fock}). Thus, the contribution
from colour octet gluon fragmentation starts at order
$\alpha_s v^7$. The short-distance coefficient was computed in
\cite{BRA94} to leading order in $\alpha_s$ and the fragmentation
function is given by
\beq
\label{lead}
D_{g\rightarrow \psi^{\prime}}(z,2m_c)={\pi \alpha_s(2m_c)\over{24
m_c^3}}
\delta(1-z)\,\langle {\cal O}^{\psi^\prime}_8(^3\!S_1)\rangle
\eeq
with $\langle {\cal O}^{\psi^\prime}_8(^3\!S_1)\rangle$
defined as in \cite{bbl}.
Braaten and Fleming found \cite{bf} that the prompt $\psi^\prime$
production cross section at the Tevatron and its $p_T$-dependence
can be fitted by $\langle
{\cal O}^{\psi^\prime}_8(^3\!S_1)\rangle = 0.0042\,
{\rm GeV}^3$, a value consistent with a $v^4$-suppression relative
to  $\langle {\cal O}^{\psi^\prime}_1(^3\!S_1)\rangle = 0.11\,
{\rm GeV}^3$.

Since the $1/p_T^4$-dependence of the production cross section is
rather generic for any fragmentation mechanism, additional
information is necessary to establish the colour octet mechanism.
One important signature of the colour octet mechanism is the  $\psi^\prime$
polarization. The polarization
of the quarkonium is calculable to a certain degree, because
the non-relativistic dynamics that governs the long-distance
evolution conserves the heavy quark spin up to corrections of order
$v^2$ that enter through the magnetic coupling $\sigma\cdot B$ in the
NRQCD Lagrangian. To implement the implications of heavy quark
spin symmetry on the tensor decomposition of matrix elements, it
is useful to represent the quarkonium state as a product of a spin
and orbital angular momentum part, similar to the decomposition in
heavy and light degrees of freedom for heavy-light mesons \cite{FAL92}.
For the $^3\!S_1^{(8)}$-operators one obtains
\beq\label{matrix1}
\langle\chi^\dagger\sigma^a T^A\,\psi\,{a_H^{(\lambda)}}^\dagger
a_H^{(\lambda)}\,\psi^\dagger\sigma^b T^A\,\chi\rangle =
\left\{\begin{array}{l}
\frac{1}{3}\,\langle {\cal O}^{\psi^\prime}_8(^3\!S_1)\rangle
\,{\epsilon^a}^*(\lambda) \epsilon^b(\lambda) \\[0.3cm]
\langle {\cal O}^{\chi_{c0}}_8(^3\!S_1)\rangle \,
\sum_{\lambda'} {\epsilon^a}^*(\lambda') \epsilon^b(\lambda')\,
|\langle 1 (\lambda-\lambda');1\lambda'| J\lambda\rangle|^2
\end{array}\right.
\eeq
where $\epsilon^a(\lambda)$ is a spin-1 polarization vector
in the quarkonium rest frame. The first line holds for
$H=\psi^\prime$ and the second for $H=\chi_{cJ}$. When inserted in
(\ref{factorization}), these equations determine the required
projections of the short-distance amplitude
on the spin (and, in general, orbital angular momentum)
of the heavy quark pair with appropriate Clebsch-Gordon coefficients.
They are equivalent to and reproduce the spin symmetry
constraints investigated in \cite{cw}. For a $\psi^\prime$ with
polarization $\lambda$, one must project on a quark pair with
equal polarization, as expected, since the two E1 transitions do
not change the spin. Since the quark pair in Fig.~\ref{diagplot}a is
transversely polarized, the projection with $\epsilon^a(0)$ vanishes,
and, at leading order in $\alpha_s$, we get
\beq\label{leadlong}
D_{g\rightarrow \psi^{\prime}_L}(z,\mu) = 0\qquad
D_{g\rightarrow \psi^{\prime}_T}(z,\mu) =
D_{g\rightarrow \psi^{\prime}}(z,\mu)
\eeq
for the fragmentation functions into longitudinally and transversely
polarized $\psi^\prime$'s. Because spin-symmetry
is not exact, long-distance
corrections of order $v^4$ arise, if the quark pair in the octet state
undergoes a double magnetic dipole transition.\\

\noindent
{\bf 4. Short-distance corrections to transverse polarization}\\[0.3cm]
\indent
Corrections to pure transverse polarization also arise, when the
polarization of the quark pair prior to nonperturbative evolution is
modified by the emission of hard gluons. The corresponding real and
(some of the) virtual corrections to gluon fragmentation are
shown in Fig.~\ref{diagplot}c and d. Let us define
\beq\label{transverseratio}
\xi \equiv 1-\frac{L}{T+L} \equiv 1-\delta \xi
\eeq
as the ratio of the (prompt) production cross section for transversely
and unpolarized $\psi^\prime$'s. Since $L$ vanishes at leading order,
$\delta\xi$  is obtained from the fragmentation function
into longitudinally polarized $\psi^\prime$'s
at order $\alpha_s^2$, while the total production rate $L+T$ in the
denominator can be evaluated in leading order approximation.

For the same reason as in leading order, the virtual corrections do not
yield longitudinally polarized $\psi^\prime$'s. This results in
considerable simplification, since along with the virtual corrections,
radiative corrections to the matrix elements in NRQCD would also have
to be calculated. The full matching calculation for the next-to-leading
short-distance coefficient of
$\langle {\cal O}^{\psi^\prime}_8(^3\!S_1)\rangle$
has been performed by Ma \cite{ma}
for unpolarized quarkonia. When combined with the result for the
fragmentation function into $\psi^\prime_L$ in eq.~(\ref{3s1}) below, the
production rates for longitudinally and transversely polarized
$\psi^\prime$'s through a quark pair in a $^3\!S_1^{(8)}$ state could be
obtained separately at next-to-leading order in $\alpha_s$.
Here we consider only the ratio $\xi$, so that the total rate
can be evaluated in leading order.

After hard gluon emission the quark pair that enters the shaded ovals
in Fig.~\ref{diagplot}c can be in various spin and angular momentum
states. At order $\alpha_s^2 v^7$, the three distinct possibilities are
$^3\!S_1^{(8)}$,$^3\!P_J^{(8)}$ and $^1\!S_0^{(8)}$. The power $v^7$
arises as combination of factors of $v$ associated with the probability
for binding a quark pair in a certain angular momentum state ($v^3$ for
$S$-waves and $v^5$ for $P$-waves) and additional factors associated
with the amplitude of the relevant Fock state in (\ref{fock}). For
$^3\!S_1^{(8)}$ and $^1\!S_0^{(8)}$ a factor $v^4$ is obtained in the
amplitude squared for the double E1 and single magnetic dipole transition,
respectively. In case of $^3\!P_J^{(8)}$ a single dipole transition
provides $v^2$.

Since there is no interference between $S$-wave and $P$-wave amplitudes
and spin-0 and spin-1 amplitudes, the contributions from the three
different intermediate quark pair states can be determined separately.
Let us turn first to the $^3\!S_1^{(8)}$ intermediate state. This is
the only one to which diagram (ciii) contributes. The projections onto
the short-distance amplitude are determined by (\ref{matrix1}). In the
actual calculation it is useful to boost into a frame where the
`+-component' of the quarkonium momentum is large and to use the
covariant projection operators of \cite{GUB80}. We then follow the
method of \cite{BRA93,BRA94} for computing fragmentation functions. The
result for the integral of the fragmentation function can be expressed
as
\beq
\int\limits_0^1 dz\,d(z,\mu,\mu') =
\int\limits_{s_{\rm min}(\mu')}^{\mu^2}\frac{d s}{s^2}
\int\limits_{4 m_c^2/s}^1 d z\,A(s,z) ,
\eeq
with $A(s,z)$ a certain projection of the amplitude squared. After
interchange of integrations, one can read off the fragmentation function.
In general, the above expression is divergent. Divergences come from
the upper limit of integration over invariant mass $s$ of the
fragmenting gluon (cut off by the fragmentation scale $\mu$ above)
due to the collinear approximation implicit in the definition of
the fragmentation function. Furthermore,
the integral over $z$  may diverge due  to soft gluon emission.

For the unpolarized  $\psi'$ fragmentation function the
logarithmically divergent part of the integration over $s$ gives
\beq\label{coldiv}
D_{g\rightarrow \psi^{\prime}}^{^3\!S_1^{(8)}}(z,\mu,\mu')
= \frac{\pi\alpha_s}{24}\,\langle {\cal O}^{\psi^\prime}_8(^3\!S_1)
\rangle\cdot\frac{\alpha_s}{2\pi} P_{g\to gg}(z) \ln\frac{{\mu}^2}
{4 m_c^2},
\eeq
where $P_{g\to gg}(z)$ is the Altarelli-Parisi gluon-gluon splitting
function. The logarithmic divergence occurs as part of the
production amplitude, when
the fragmenting gluon splits into two collinear gluons of low
invariant mass. Since both gluons are parallel and transverse, so is
the final quark pair. Thus, the $s$-integral is finite for
longitudinally polarized $\psi^\prime$'s.

For a quark pair in a $^3\!S_1^{(8)}$ state the integral over $z$
is finite. A divergence can potentially arise for $P$-waves, when the
amplitude is expanded in the relative velocity. However, soft gluon
emission with energy less than $\mu'$, the NRQCD factorization scale,
should be considered as part of a NRQCD matrix element. For
longitudinal polarization there is no contribution at order $\alpha_s$,
into which an infrared logarithm could be absorbed and all longitudinal
fragmentation functions must be finite as $z\to 1$. The integral
above can then be evaluated  taking $\mu$  to infinity and $\mu'$
to zero in all cases considered  in this paper.

The result for the fragmentation function into
longitudinally polarized $\psi^\prime$'s
through a colour octet $^3\!S_1$ quark pair is
\beq\label{3s1}
D_{g\rightarrow \psi^{\prime}_L}^{^3\!S_1^{(8)}}(z,2 m_c)
= \frac{\alpha_s^2(2 m_c)}{24} N_c\,\frac{1-z}{z}\,
\frac{\langle {\cal O}^{\psi^\prime}_8(^3\!S_1)\rangle}{m_c^3} ,
\eeq
where $N_c=3$ is the number of colours.

For a $\psi^\prime$ produced through an intermediate quark pair with
orbital angular momentum one, we need the matrix elements
\beq\label{matrix2}
\langle\chi^\dagger\sigma^a T^A\left(-\frac{i}{2}
\stackrel{\leftrightarrow}{D}_i\right)
\psi\,{a_H^{(\lambda)}}^\dagger
a_H^{(\lambda)}\,\psi^\dagger\sigma^b T^A\left(-\frac{i}{2}
\stackrel{\leftrightarrow}{D}_j\right)\chi\rangle =
\langle {\cal O}^{\psi^\prime}_8(^3\!P_0)\rangle\,
\delta_{ij}\,{\epsilon^a}^*(\lambda)\epsilon^b(\lambda) .
\eeq
The right hand side incorporates the constraints
from spin symmetry and is accurate up to
$v^2$-corrections. Thus, we have to project the short-distance amplitude
on a longitudinally polarized quark pair and sum over all orbital
angular momentum states. Note that the total angular momentum $J$ of the
quark pair is not specified. The resulting fragmentation function is
\beqa\label{3pj}
D_{g\rightarrow \psi^{\prime}_L}^{^3\!P_J^{(8)}}(z,2 m_c)
&=& \frac{\alpha_s^2(2 m_c)}{16} \frac{N_c^2-4}{N_c}
\Bigg[\frac{z^2+8 z-12}{z^2} (1-z)\ln (1-z)
\nonumber\\
&&\,- \frac{2 z^3+z^2-28 z+24}{2 z}\Bigg]
\frac{\langle {\cal O}^{\psi^\prime}_8(^3\!P_0)\rangle}{m_c^5} .
\eeqa
Because of the decomposition of
the matrix element (\ref{matrix2}), which would differ
for $\chi_{cJ}$ states, this fragmentation function
can not be obtained as a linear
combination of the short-distance part of the
fragmentation functions into polarized
$\chi_{cJ}$ states \cite{ctw}. Note also that it is finite as $z\to 1$,
as expected from the considerations above.

Finally we have to account for the quark pair in a spin-0 state. While
in the previous two cases, the nonperturbative evolution could be
considered as spin-conserving at
order $v^7$, a non-zero value for
$\langle {\cal O}^{\psi^\prime}_8(^1\!S_0)\rangle$ arises through a
magnetic dipole transition at order $v^7$. Since the initial quark pair
evolves into a $\psi^\prime$ with a probability independent of the
polarization of $\psi^\prime$, the fragmentation function for a
longitudinally polarized $\psi^\prime$ is given by dividing the total
fragmentation probability by three. The result is
\beq\label{1s0}
D_{g\rightarrow \psi^{\prime}_L}^{^1\!S_0^{(8)}}(z,2 m_c)
= \frac{\alpha_s^2(2 m_c)}{864} \frac{N_c^2-4}{N_c}
\left[3 z-2 z^2+2 (1-z)\ln (1-z)\right]
\frac{\langle {\cal O}^{\psi^\prime}_8(^1\!S_0)\rangle}{m_c^3} .
\eeq
The functional dependence is the same as for the colour singlet
fragmentation function for $\eta_c$ in \cite{BRA93}.

The fragmentation function $D_{g\rightarrow \psi^{\prime}_L}(z)$
for longitudinally polarized $\psi^\prime$'s
is given by the sum of (\ref{3s1}), (\ref{3pj}) and (\ref{1s0}) at
leading non-vanishing order in $\alpha_s$. There is also a contribution
from charm fragmentation, which is formally of the same order
$\alpha_s^2 v^7$, where the quark pair in Fig.~\ref{diagplot}b is in
an octet state. The corresponding fragmentation functions are identical
up to a colour factor to the colour singlet charm fragmentation functions,
which contribute at order $\alpha_s^2 v^3$.
Since colour singlet fragmentation (including charm fragmentation)
can account only for $1/30$ of all $\psi^\prime$'s, colour octet
charm fragmentation contributes at most a tiny fraction
to the total production rate. It can be neglected without affecting
our conclusions.\\

\noindent
{\bf 5. Results}\\[0.3cm]
\indent
The production cross section for longitudinally polarized $\psi^\prime$'s
at the Tevatron is obtained by convoluting the fragmentation
function with the gluon production cross section in $p\bar{p}$ collisions
with center-of mass-energy $\sqrt{s}=1.8\,{\rm TeV}$,
\beq\label{convolve}
\frac{d\sigma}{d p_T}(p\bar{p}\to \psi^\prime_L + X) =
\int\limits_{2p_T/\sqrt{s}}^1 d z\,\frac{d\sigma}{d p_T}(p\bar{p}\to
g(P_T/z) + X)\,D_{g\rightarrow \psi^{\prime}_L}(z,\mu) .
\eeq
The longitudinally polarized production fraction $\delta\xi$ is given by
dividing this expression by a similar expression with
$D_{g\rightarrow \psi^{\prime}_L}(z)$ replaced by
$D_{g\rightarrow \psi^{\prime}}(z)$ given in (\ref{lead}). We have used the
CTEQ3L parton distributions to evaluate the gluon production cross
section. The dominant production channel is gluon-gluon fusion, while
the $q\bar{q}$ channel is negligible. The gluon fragmentation functions
are evolved from the initial scale $2 m_c= 3\,{\rm GeV}$ to $\mu$ of
order $p_T$ by the usual Altarelli-Parisi equations. To very good
approximation, mixing can be neglected in solving the AP equations.
As input we have chosen $\alpha_s(2 m_c)=0.27$.
In addition we have implemented the pseudo-rapidity cut $|\eta|>0.6$ as
in the latest CDF analysis \cite{cdf}.
\begin{table}[t]
\addtolength{\arraycolsep}{0.2cm}
$$
\begin{array}{|c||c|c|c|c|}
\hline
p_T/{\rm GeV} & r_1 & r_2 & r_3 & \delta\xi_{SD}\\ \hline
6 & 0.027 & 0.035 & 0.001 & 0.062 \\
9 & 0.022 & 0.030 & 0.001 & 0.052 \\
12 & 0.020 & 0.027 & 0.001 & 0.047 \\
15 & 0.018 & 0.025 & 0.001 & 0.043 \\
18 & 0.017 & 0.024 & 0.001 & 0.041 \\
21 & 0.015 & 0.024 & 0.001 & 0.039 \\
24 & 0.015 & 0.023 & 0.001 & 0.038 \\
\hline
\end{array}
$$
\caption{\label{table} The coefficients $r_i$ as a function of $p_T$ with
$\mu=p_T$ and $|\eta|\!>\!0.6$. The last column gives
the deviation from pure transverse polarization,
$\delta \xi_{SD}=r_1+r_2$,
for $\langle {\cal O}^{\psi^\prime}_8(^3\!P_0)\rangle/
\langle {\cal O}^{\psi^\prime}_8(^3\!S_1)\rangle =1$ and neglecting the
contribution from $\langle {\cal O}^{\psi^\prime}_8(^1\!S_0)\rangle$.}
\end{table}

The final result for $\delta \xi$ can be represented as the sum of
the three contributions to longitudinal fragmentation discussed
earlier:
\beq\label{rs}
\delta\xi_{SD}(p_T,\mu) =
r_1(p_T,\mu)+r_2(p_T,\mu) \frac{1}{m_c^2}
\frac{\langle {\cal O}^{\psi^\prime}_8(^3\!P_0)
\rangle}{\langle {\cal O}^{\psi^\prime}_8(^3\!S_1)\rangle} +
r_3(p_T,\mu) \frac{\langle {\cal O}^{\psi^\prime}_8(^1\!S_0)\rangle}
{\langle {\cal O}^{\psi^\prime}_8(^3\!S_1)\rangle} .
\eeq
We have added a subscript `SD' to indicate that spin symmetry breaking
long distance corrections of order $\alpha_s v^{11}$
are not included here. The coefficients $r_i$ are listed in
Table~\ref{table} for $\mu=p_T$.

As can be seen from the Table, the contribution from $r_3$ is small and
can be dropped for any reasonably expected ratio of the matrix elements
that multiplies $r_3$. This leaves us with one ratio of matrix elements
which is not very well constrained at present. The velocity scaling rules
tell us that this ratio should be of order unity. A slightly more
definite, but crude estimate can be obtained by the following argument
borrowed from \cite{BOD92}: The operators ${\cal O}^{\psi^\prime}_8
(^3\!P_0)$ and ${\cal O}^{\psi^\prime}_8(^3\!S_1)$ mix under
renormalization \cite{bbl}. The solution to the renormalization
group equation in the one-loop approximation is given by
\beq\label{rge}
\langle {\cal O}^{\psi^\prime}_8(^3\!S_1)\rangle(\mu') =
\langle {\cal O}^{\psi^\prime}_8(^3\!S_1)\rangle(\mu_0^\prime) +
\frac{6 (N_c^2-4)}{N_c\beta_0 m_c^2} \ln\frac{\alpha_s(\mu_0^\prime)}
{\alpha_s(\mu')}\,\langle {\cal O}^{\psi^\prime}_8(^3\!P_0)\rangle ,
\eeq
where we used $\langle {\cal O}^{\psi^\prime}_8(^3\!P_J)\rangle =
(2 J+1)\,\langle {\cal O}^{\psi^\prime}_8(^3\!P_0)\rangle$, as follows
from (\ref{matrix2}), and $\beta_0=11N_c/6-N_f/3$. The second
term is formally enhanced by a logarithm if $\mu_0'$ and $\mu'$
are very different. Neglecting the first term on the right
hand side and choosing
the initial scale of order of the typical non-relativistic momenta
$\mu_0'\approx m_c v$ and $\mu'\approx m_c$, we get,
with $\alpha_s(1.5\,{\rm GeV})=0.35$, the estimate
\beq
\frac{1}{m_c^2}\frac{\langle {\cal O}^{\psi^\prime}_8(^3\!P_0)
\rangle}{\langle {\cal O}^{\psi^\prime}_8(^3\!S_1)\rangle}
\approx 1.25 .
\eeq
\begin{figure}[t]
   \vspace{-1cm}
   \centerline{\epsffile{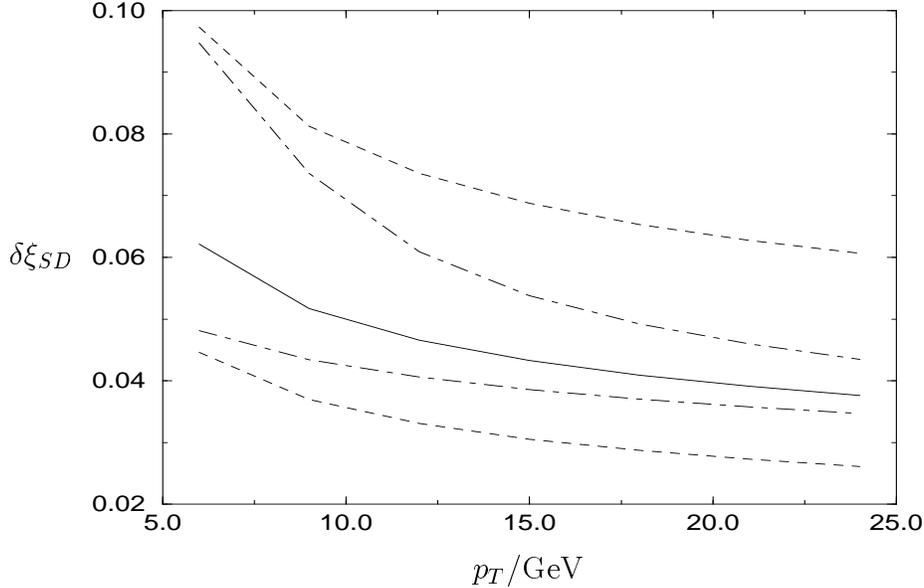}}
   \vspace*{0cm}
\caption{\label{graph} The short-distance correction $\delta\xi_{SD}$
as function of $p_T$. The solid line shows the prediction for $\mu=p_T$
and $\langle {\cal O}^{\psi^\prime}_8(^3\!P_0)\rangle/
\langle {\cal O}^{\psi^\prime}_8(^3\!S_1)\rangle =1$. The two dash-dotted
lines are for $\mu=p_T/2$ (upper curve) and $\mu=2 p_T$ (lower
curve) at fixed $\langle {\cal O}^{\psi^\prime}_8(^3\!P_0)\rangle/
\langle {\cal O}^{\psi^\prime}_8(^3\!S_1)\rangle =1$. The two dashed
lines correspond to variation of
$\langle {\cal O}^{\psi^\prime}_8(^3\!P_0)\rangle/
\langle {\cal O}^{\psi^\prime}_8(^3\!S_1)\rangle$ between 0.5 and 2 at
fixed $\mu=p_T$.}
\end{figure}
Although the logarithm is in fact not large, a comparable estimate
for $\chi_c$ states as in \cite{BOD92} is only off by a factor of two from
the phenomenologically known ratio in that case. If we assume that this
difference arises from an inaccurate choice of the initial scale and
adjust the logarithm to the known ratio for matrix elements in the
$\chi_c$ case, the estimate above changes from 1.25 to 0.65. In view of
this we consider a ratio of one as our best guess. With this choice
the prediction for $\delta\xi_{SD}$ is shown in the last column of the
Table and as solid line in Fig.~\ref{graph}. The Figure also shows
the prediction if we modify the above ratio of matrix elements to $2$
or to $1/2$. This uncertainty constitutes the dominant source of
theoretical error until $\langle {\cal O}^{\psi^\prime}_8(^3\!P_0)
\rangle$ is determined from another process. For comparison, we have
plotted the variation of $\delta\xi$, when the renormalization scale
is set to $\mu=p_T/2$ or $\mu=2 p_T$. This uncertainty is about
$\pm 20\%$, except at small $p_T$, where the choice $p_T/2$
is not very well motivated. It should also be mentioned that by nature
of the fragmentation approximation, the prediction is more reliable
at large $p_T$. If one integrates the production cross section over
$p_T$ starting from some $p_{T,{\rm min}}$, the corresponding integrated
$\delta\xi$ is roughly equal to $\delta\xi$ at $p_{T,{\rm min}}$, because
the absolute production cross section decreases rapidly with $p_T$.\\

\noindent
{\bf 6. Discussion}\\[0.3cm]
\indent
{}From the above we conclude that short-distance corrections to the
fragmentation functions decrease the transverse polarization of
100\% at leading order in $\alpha_s$ by approximately $5\%$ and
hardly more than $10\%$, even if the $^3\!P_0$-matrix element is
large. We have ignored the colour singlet production mechanisms,
since they account for only $3\%$ of all $\psi^\prime$'s. If we
assume that the singlet mechanisms produce
randomly polarized $\psi^\prime$'s,
they would contribute only an additional percent to the longitudinal
polarization fraction. A potential correction of order $4 m_c^2/p_T^2$
comes from the fragmentation approximation. However, the octet
production processes at leading order in $\alpha_s$ yield basically
pure transverse polarization even without the fragmentation
approximation to the matrix elements due to a particular
structure of the gluon-gluon fusion amplitude \cite{cl}. Thus the
higher twist correction is suppressed by $(\alpha_s/\pi)\,4 m_c^2/p_T^2$,
which is small for sufficiently large $p_T\sim 10\,{\rm GeV}$.

Deviations from pure
transverse polarization arise from spin symmetry breaking already at
leading order in $\alpha_s$, when the quark pair in Fig.~\ref{diagplot}a
undergoes  spin-changing nonperturbative evolution.
Because of parity and charge conjugation symmetry these corrections scale
with $v^4$.
Since random polarization would give $\delta\xi=1/3$,
a natural estimate of the long-distance correction is
$\delta\xi_{LD}={\cal O}(v^4/3) \approx (3-4)\%$.
Despite their smallness, the short distance corrections
constitute the dominant source of depolarization.
We estimate, conservatively, that if the
colour octet mechanism does account for the observed prompt $\psi^\prime$
production cross section,  then  the $\psi^\prime$'s should be
$(85-95)\%$ transversely polarized.

This prediction can be tested rather simply by measuring
the lepton angular distribution in the $\psi^\prime
\rightarrow l^+l^-$ decay channel,
\beq
\frac{d\Gamma}{d\cos\theta}\propto 1+
\frac{1-3\delta\xi}{1+\delta\xi}\,\cos^2 \theta,
\eeq
where $\theta$ denotes the angle between the lepton three momentum in
the $\psi^\prime$ rest frame and the $\psi^\prime$ three momentum in
the lab frame. As statistics improves, a  measurement of the
angular distribution will become feasible. Time will tell whether
such a measurement, together with other signatures of the colour octet
mechanism, such as those in $e^+ e^-$ annihilation \cite{sonst},
provides a consistent picture of quarkonium production that includes
the  relativistic structure of the bound state.\\

\noindent
{\bf Acknowledgements.} We benefitted from conversations
with Wolfgang Kilian and Mark Wise.  M.~B. is supported by the
Alexander von Humboldt foundation.

\end{document}